\documentclass[english,aps,prl,superscriptaddress,floatfix,notitlepage,reprint,show pacs]{revtex4-2}
\usepackage[T1]{fontenc}
\usepackage[latin9]{inputenc}
\usepackage{physics}
\usepackage{natbib}
\usepackage{amsthm}
\usepackage{amssymb}
\usepackage{dsfont}
\usepackage{amsmath}
\usepackage{bm}
\makeatletter
\theoremstyle{plain}

\theoremstyle{plain}

\theoremstyle{plain}
\newtheorem*{prop*}{\protect\propositionname}

\usepackage{braket}
\usepackage{txfonts} 
\usepackage{graphicx}
\usepackage[usenames,dvipsnames]{xcolor}
\usepackage[colorlinks=true,citecolor=Blue,linkcolor=RubineRed,urlcolor=Blue]{hyperref}
\usepackage{tikz}
\usepackage{rotating} 
\date{\today}

\def\k{\kappa}

\newcommand{\beq}{\begin{equation}}
\newcommand{\eeq}{\end{equation}}
\newcommand{\beqa}{\begin{eqnarray}}
\newcommand{\eeqa}{\end{eqnarray}}

\makeatother

\begin{document}

\title{Periodic dynamics of population-imbalanced fermionic condensates in optical lattices}
\author{Avinaba Mukherjee}
\email{avinaba.mukherjee@rediffmail.com}
\author{Raka Dasgupta}
\email{rdphy@caluniv.ac.in}
\address{Department of Physics, University of Calcutta, $92$ A. P. C. Road, Kolkata $700009$, India}

\begin{abstract}
 We investigate the dynamics of a population-imbalanced two-species fermionic system trapped in an optical lattice. The paired fermions here can form bosonic molecules via Feshbach coupling in the presence of an external magnetic field.  It is shown that the natural fluctuations of the condensate fraction are periodic beyond a threshold Feshbach detuning; and below this threshold value, the condensate fraction shows no oscillation at all.  The oscillation frequency vs. detuning curve is linear in nature. The slope and intercept of this line are shown to carry important information about the amount of imbalance present in the system, and the momentum space structure of the exotic phases. 
\end{abstract}
\maketitle
\section{Introduction}
Population-imbalanced ultracold fermionic systems remain an active field of research for the past two decades. While it is known that Feshbach-coupled two-species fermionic gases can demonstrate a BCS-BEC crossover as the effective interaction is varied \cite{BCS-BEC1,BCS-BEC2,BCS-BEC3,BCS-BEC4,BCS-BEC5}, the situation changes if the system resembles a spin-polarized state i.e., one species is more populated than the other. In this case, instead of the weakly interacting homogeneous BCS state (that demands both species to be equally populated), one would have novel pairing structures that can accommodate the unpaired fermions.  Several theoretically proposed phases exist for such systems, including (i) Breached pair (BP) \cite{bp_reference1,bp_reference2,bp_reference3,bp_reference4,bp_reference5,bp_reference6,bp_new2,bp_another1,bp_another2} (ii) Phase separation (PS) \cite{ps_reference1,ps_reference2,ps_reference3} and (iii) Fulde-Ferrell-Larkin-Ovchinnikov (FFLO) states \cite{FFLO_reference1,FFLO_reference2,FFLO_reference3, max_planck,FFLO_new1,FFLO_new2}. A new phase with a non-FFLO spatially modulated pairing state has also been reported recently \cite{new_phase}. 

The tunability associated with the ultracold atomic systems went a step ahead with the advent of optical lattices \cite{optical_lattice_reference1,optical_lattice_reference2,optical_lattice_reference3,optical_lattice_reference4} as now the geometry as well as the dimension of the lattice could be easily controlled by using the laser beams. As fermionic superfluidity could now be realized in the optical lattices \cite{BCS-BEC5,fermionic_superfluid,hopping6}, it opened up the possibility of studying the non-BCS exotic pairing phases in the presence of the lattice potentials\cite{sarma1, sarma3, ps3, sarma4, fflo1, fflo3, ps2, ps4, ps, ps1, fflo2, fflo_reference4, fflo_reference5, fflo_reference6}. However, it is extremely difficult to find direct experimental signatures of the theoretically predicted exotic phases (except the PS state \cite{ps_reference3}), and a few indirect methods has been proposed in the past \cite{ alt, lus, pec, anna, singh, bp_diagram, kaj, zhou, bp_new2,continuous}.\\

The study of the out-of-equilibrium dynamics and fluctuation dynamics of ultracold atoms is both an interesting and challenging field \cite{superexchange}, because of the in-built nonlinearity of the systems.  Oscillatory dynamics of atoms in optical lattices have been probed in the context of Bloch oscillations \cite{bloch_osc, bloch_osc2}, Collective oscillations \cite{goldstone_mode_BEC,goldstone_mode_BEC2,goldstone_mode_BEC3,higgs_mode_BEC,higgs_mode_BEC2,higgs_mode_BEC3,higgs_mode_BEC4,bogoliubov_mode_BEC,bogoliubov_mode_BEC2,collective_mode_BEC_BCS}, Josephson junction arrays \cite{josephson} and spin imbalance dynamics in the presence of superexchange interactions \cite{superexchange, superexchange2}. Most of these involve population oscillations between either two quantum states or two bands. Quantum transport-related properties of ultracold atoms in optical lattices have been studied as well \cite{transp1,transp2,transp3,transp4,transp5,transp6}, where the system is intrinsically non-equilibrium, with almost no equilibrium counterpart being present.  In this work, however, we start from the equilibrium configurations of the population-imbalanced systems, and explore the fluctuation dynamics on top of it. It is observed that an oscillatory pattern arises in these fluctuations, even at the mean-field level.

In this paper, we study the mean-field level fluctuation dynamics of the population-imbalanced fermionic condensate in an optical lattice for 1D, 2D and 3D. We observe that the dynamics is periodic, and the oscillation frequency changes almost linearly with a varying Feshbach detuning. We can obtain information about the amount of population imbalance present in the system, and also the nature of the pairing, from the slope and intercept of this straight line.

The paper is organized as follows. The basic theoretical model and the dynamical equations are described in Sec. \ref{2nd section}. In Sec. \ref{3rd section},  possible pairing phases are discussed and the relevant parameters are set up in different dimensions. The frequency vs. detuning plots are presented in Sec. \ref{4th section}, and it is shown that the oscillation dynamics are periodic. How the intercept of the frequency vs. detuning curve varies with the amount of population imbalance is discussed in Sec. \ref{6th section}, and the same for the slope is done in Sec. \ref{5th section}. In Sec. \ref{7th section}, we provide the analytical reasoning for the numerical results of the previous two sections. The results are summarized in Sec. \ref{8th section}.

\section{Model Hamiltonian and dynamics}\label{2nd section}
Here we consider a system of ultracold fermionic atoms in an optical lattice, that can form a bosonic molecule by means of Feshbach resonances. The system can be described by an extended version of the Hubbard Hamiltonian
\begin{equation}
\begin{split}
H= & -t \sum_{\langle \vec j \vec l\rangle}  (a^\dagger_{\vec j\uparrow} a_{\vec l\uparrow}+a^\dagger_{\vec j\downarrow} a_{\vec l\downarrow}) -\sum_{ \vec j} (\mu_\uparrow a^\dagger_{\vec j\uparrow} a_{\vec j\uparrow}+\mu_\downarrow a^\dagger_{\vec j\downarrow} a_{\vec j\downarrow})\\ & +g_1 \sum_{\vec j}{n_{\vec j\uparrow}n_{\vec j\downarrow}} +g_2\Big( \sum_{\vec j} b^\dagger_{\vec j}  a_{\vec j\downarrow} a_{\vec j\uparrow} +\sum_{\vec j} a^\dagger_{\vec j\uparrow} a^\dagger_{\vec j\downarrow} b_{\vec j}\Big)\\
& +[\epsilon_b-(\mu_\uparrow+\mu_ \downarrow)]\sum_{\vec j}{b^\dagger_{\vec j}b_{\vec j}}
\end{split}
\end{equation}
Here, up-spin $(\uparrow)$  and down-spin $(\downarrow)$  represent either two different atomic species or two different hyperfine states of the same atom; $\mu_{\uparrow}$ and $\mu_{\downarrow}$ are their respective chemical potentials.  The hopping amplitude  (assumed to be the same for both species) is $t$. Also, $a^\dagger_{\vec j}$ and $a_{\vec j}$ are respectively the fermionic atom creation and annihilation operators for site $\vec j =(j_x, j_y, j_z)$, and their bosonic counterparts for molecule creation and destruction are given by $b^\dagger_{\vec j}$ and $b_{\vec j}$. The number operators for up-spin and down-spin fermions at $\vec j$ th site are $n_{\Vec{j}\uparrow}$ and  $n_{\Vec{j}\downarrow}$ respectively. The on-site interaction strength is $g_1$, while $g_2$ is the additional interaction strength of the Feshbach variety which couples two fermionic atoms to form a bosonic molecule. The threshold energy of the composite bosonic molecule energy band is $\epsilon_b$, and it is the pivotal adjustable ``detuning" parameter in this model.

This Hamiltonian takes the following form in momentum space : 
\begin{equation}
\begin{split}
 H=& \sum_{\vec k}\epsilon_{\vec k\uparrow}a^\dagger_{\vec k\uparrow} a_{\vec k\uparrow}+\sum_{\vec k}\epsilon_{-\vec k\downarrow}a^\dagger_{-\vec k\downarrow} a_{-\vec k\downarrow}
\\ &+g_1\sum_{\vec k,\vec k'} a^\dagger_{\vec k\uparrow} a^\dagger_{-\vec k+\vec q\downarrow} a_{-\vec k'+\vec q \downarrow} a_{\vec k'\uparrow}\\  &+g_2(\sum_{\vec q,\vec k} b^\dagger_{\vec q} a_{-\vec k+\vec q \downarrow} a_{\vec k\uparrow}+\sum_{\vec k,\vec q}a^\dagger_{\vec k\uparrow} a^\dagger_{-\vec k+\vec q\downarrow}b_{\vec q})\\
 & +[\epsilon_b-(\mu_\uparrow+\mu_ \downarrow)]\sum_{\vec q} b^\dagger_{\vec q} b_{\vec q}
 \end{split}
\end{equation}
$\epsilon_{\Vec{k}\uparrow}=-2t(\cos{\vec{k}\vec {a}})-\mu_{\uparrow}$ and $\epsilon_{-\Vec{k}\downarrow}=-2t(\cos{\vec {k}\vec {a}})-\mu_{\downarrow}$ are the energy-momentum dispersion relation for up-spin and down-spin respectively. Here, $\vec{a}$ is the lattice vector of the optical lattice. Also, $a^\dagger_{\vec k\uparrow}$, $a_{\vec k\uparrow}$ are the creation and annihilation operators corresponding to momentum $k$ for one fermionic species, and $a^\dagger_{\vec k \downarrow}$, $a_{\vec k \downarrow}$ are the corresponding operators for the other fermionic species. The bosonic molecule creation and annihilation operators in the momentum space are given by  $b^\dagger_{\vec q}$ and $b_{\vec q}$ respectively. Here $\vec q$ is the momentum of the composite bosons and has a magnitude zero for BCS-type of pairing, and a non-zero magnitude if an exotic pairing structure like FFLO is involved. More details about these pairing phases are discussed in \ref{different phases}.

We define 
\begin{equation}
  O_{\vec k,\vec q}=\langle a_{-\vec k+\vec q\downarrow} a_{\vec k\uparrow}\rangle 
\end{equation}
 
 The equilibrium conditions are given by 
 
 \begin{equation}
 \frac{\partial O^{eq}_{\vec{k},\vec{q}}}{\partial t}=0 \hspace{15 pt} \mbox{and} \hspace{15 pt}  \frac{\partial b^{eq}_{\vec q}}{\partial t}=0
\end{equation} 

Now, we consider the intrinsic and spontaneous quantum fluctuations on top of the equilibrium base states. We write, $O_{\vec k,\vec q}=O^{eq}_{\vec{k},\vec{q}}+\Tilde O_{\vec k,\vec q}$ and
$b_{\vec q}=b^{eq}_{\vec q} +\Tilde b_{\vec q}$.  Here, $b^{eq}_{\vec{q}}$ and $O^{eq}_{\vec{k},\vec{q}}$ are the equilibrium value of $b_{\vec q}$ and $O_{\vec k,\vec q}$ respectively.
$\Tilde O_{\vec k,\vec q}$ is the fluctuation in $O_{\vec k,\vec q}$ and $\Tilde b_{\vec q}$ is the fluctuation in $b_{\vec q}$. The scheme resembles the one used in \cite{continuous} for studying population-imbalanced uniform systems as well as trapped systems under local density approximation. Here we extend that scheme to incorporate the effects of optical lattices that are more relevant for present-day experiments. 

By calculating the commutation relations of   $b_{\vec q}$ and $O_{\vec k,\vec q}$ respectively with the Hamiltonian, and imposing the  equilibrium conditions,  we arrive at two dynamical equations : 
\begin{equation}
\begin{split}
\label{commutation of O}
i\hbar\frac{\partial\Tilde O_{\vec k,\vec q}}{\partial t}= & \sum_{\vec k}(\epsilon_{\vec k\uparrow}+\epsilon_{-\vec k+\vec q \downarrow})\Tilde O_{\vec k,\vec q}
  -(g_1\sum_{\vec k}\Tilde O_{\vec k,\vec q}+g_2 \Tilde b_{\vec q})
\end{split}
\end{equation}
\begin{equation}
\label{boson}
i\hbar\frac{\partial \Tilde b_{\vec q}}{\partial t}=g_2\sum_{\vec k}\Tilde O_{\vec k}+(\epsilon_b-(\mu_\uparrow+\mu_\downarrow))\Tilde b_{\vec q}
\end{equation}
By taking the Fourier transform of Eqs. (\ref{commutation of O}, \ref{boson}), we obtain 
\begin{equation}
\label{momentum space representation of O}
  \sum_{\Vec{k}} \Tilde O_{\vec k,\vec q}(\omega)=\frac{g_2 \Tilde b_{\vec q}(\omega) f_1(\omega)}{1-g_1 f_1(\omega)}
\end{equation}
Here 
\begin{equation}
\label{f1_omega}
 f_1(\omega)=\sum_{\Vec{k}}\frac{1}{\epsilon_{\vec k\uparrow}+\epsilon_{-\vec k+\vec q\downarrow}+\hbar \omega}
\end{equation} 

By putting   Eq. (\ref{momentum space representation of O}) in Eq. (\ref{boson}) we  get
\begin{equation}
   b_{\vec q}(\omega)\Big( \epsilon_b+\hbar \omega+\frac{g^2_2  f_1(\omega)}{1-g_1 f_1(\omega)}\Big)=0
\end{equation}

Thus, in the Fourier expansion of $b_{\vec q}(t)$ , only those $b_{\Vec{q}}(\omega)$ would survive for which 
\begin{equation}
\label{freq}
\epsilon_b+\hbar \omega+\frac{g^2_2  f_1(\omega)}{1-g_1 f_1(\omega)}=0
\end{equation}

Therefore, if there is a single solution of $\omega$ for  Eq. (\ref{freq}), then  $b_{\Vec{q}}(t)=b^{eq}_{\vec{q}}+b_1 e^{i\omega t}$ and the condensate fraction ($|b_{\vec q}(t)|^2$)  would show an oscillatory dynamics with frequency $\pm\omega$. If there are two solutions $\omega_1$ and $\omega_2$: $b_{\vec{q}}(t)=b^{eq}_{\vec{q}}+b_1 e^{i\omega_1 t}+b_2 e^{i\omega_2 t}$.
So, the condensate fraction has three periodic components: $\pm\omega_1$, $\pm\omega_2$, $\pm (\omega_1-\omega_2)$ and similarly for higher number of allowed $\omega$ values. If no real solution is found for Eq. (\ref{freq}), then the condensate fraction would have no oscillatory dynamics at all. On the BEC side, these essentially are the Bogoliubov modes of oscillation \cite{RMP}.

As evident from Eq. (\ref{f1_omega}), the quantity $f_1(\omega)$ involves a sum over all paired regions in the momentum space and is dependent on (i) the dimension of the system, (ii) the amount of population imbalance present in the system, and  (iii) the nature of pairing. Thus, the solution for $\omega$ obtained from Eq. (\ref{freq}), which translates into the frequency of oscillation of the condensate fraction, would contain information about these factors as well. 

\section {Different variants of the population imbalanced system }\label{3rd section}
\subsection{Dimension of the system} 
Most ultracold-atomic experiments offer tremendous flexibility when it comes to the dimensionality of the system. For example, a 1-dimensional (1D) optical lattice can be formed when the potential is shallower in one direction but it is much deeper in the other two directions \cite{1D1,1D2,1D3,1D4}.
Converting the summation in Eq. (\ref{f1_omega}) to an integral, 
\begin{equation}
\label{f1 1D}
  f_1(\omega)=\frac{L}{2\pi\hbar} \int \frac{dk}{\epsilon_{\vec k\uparrow}+\epsilon_{-\vec k+\vec q\downarrow}+\hbar \omega}  
\end{equation}
Here  $\epsilon_{\vec k\uparrow}=-2t(\cos{k a})-\mu_{\uparrow}$, 
$\epsilon_{-\vec k\downarrow}=-2t(\cos{k a})-\mu_{\downarrow}$. $L$ is the quantization length of the optical lattice.
Similarly, a 2-dimensional (2D) optical lattice can be formed when the potential is deeper in one direction but shallower in the other two directions \cite{2D}.
\begin{equation}
\label{f1 2D}
    f_1(\omega)=\frac{A}{(2\pi\hbar)^2} \int \frac{kdkd\theta}{\epsilon_{\vec k\uparrow}+\epsilon_{-\vec k+\vec q\downarrow}+\hbar \omega} 
\end{equation}

In this case, 
 $\epsilon_{\vec k\uparrow}=-2t(\cos{k_x a}+\cos{k_y a})-\mu_{\uparrow}$, 
 $\epsilon_{-\vec k\downarrow}=-2t(\cos{k_x a}+\cos{k_y a})-\mu_{\downarrow}$ and $A$ is the quantization area of the optical lattice.

In 3-dimensional (3D) optical lattices, the potential depths are comparable in three possible directions \cite{3D}.
Here,
\begin{equation}
\label{f1 3D}
    f_1(\omega)=\frac{V}{(2\pi\hbar)^3} \int \frac{k^2 sin{\theta} dk d\theta d\phi}{\epsilon_{\vec k\uparrow}+\epsilon_{-\vec k+\vec q\downarrow}+\hbar \omega}
\end{equation}
In this case,
$\epsilon_{\vec k\uparrow}=-2t(\cos k_x a+\cos k_y a+\cos k_z a)-\mu_\uparrow$, 
$\epsilon_{\vec k\downarrow}=-2t(\cos k_x a+\cos k_y a+\cos k_z a)-\mu_\downarrow$ and $V$ is the quantization volume of the optical lattice. In Eqs. (\ref{f1 2D}) and (\ref{f1 3D}), it is assumed that the lattice constant is the same value in all directions. 
\subsection{Possible pairing phases} 
\label{different phases}
\begin{figure}[t]
    \includegraphics[width=0.6\linewidth]{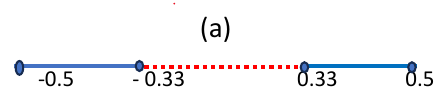}
    \includegraphics[width=0.6\linewidth]{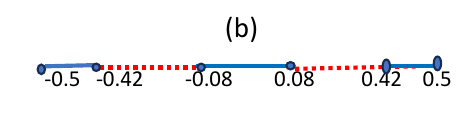}
   \includegraphics[width=0.6\linewidth]{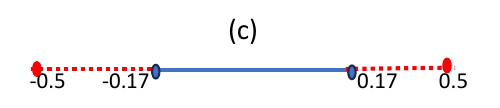}
    \includegraphics[width=0.6\linewidth]{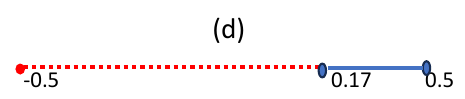}
    \caption{The structure of different pairing phases in population imbalanced fermionic systems in unit of $\pi$ for 1D of (a) BP1 (b) BP2 (c) PS (d) FFLO: Blue and red indicates paired and unpaired region at $P=0.5$}
    \label{all phases} 
\end{figure}
\begin{figure}[t]
    \includegraphics[width=0.6\linewidth]{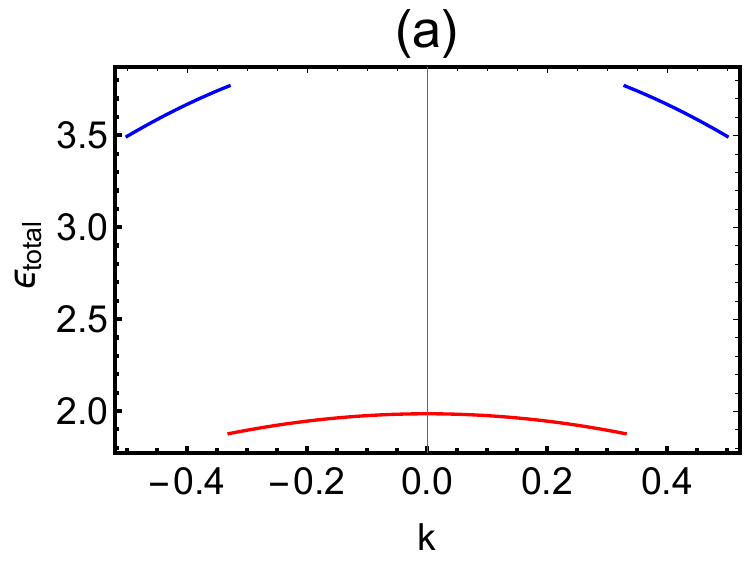}
    \includegraphics[width=0.6\linewidth]{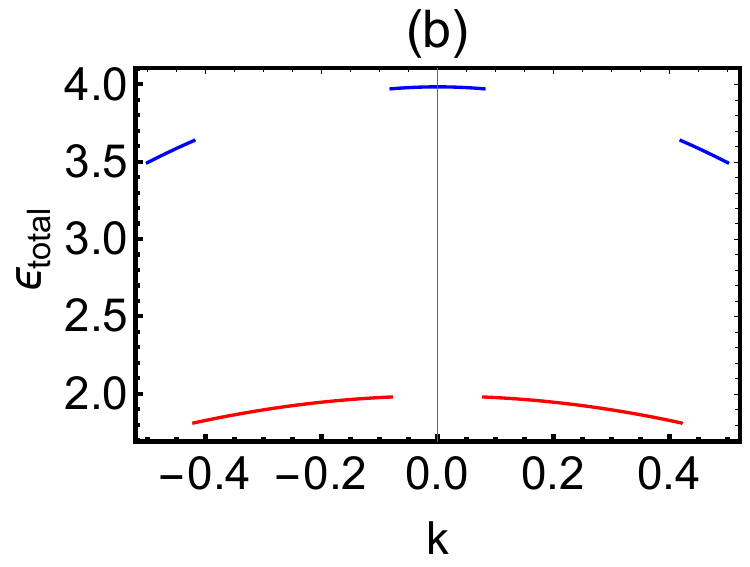}
   \includegraphics[width=0.6\linewidth]{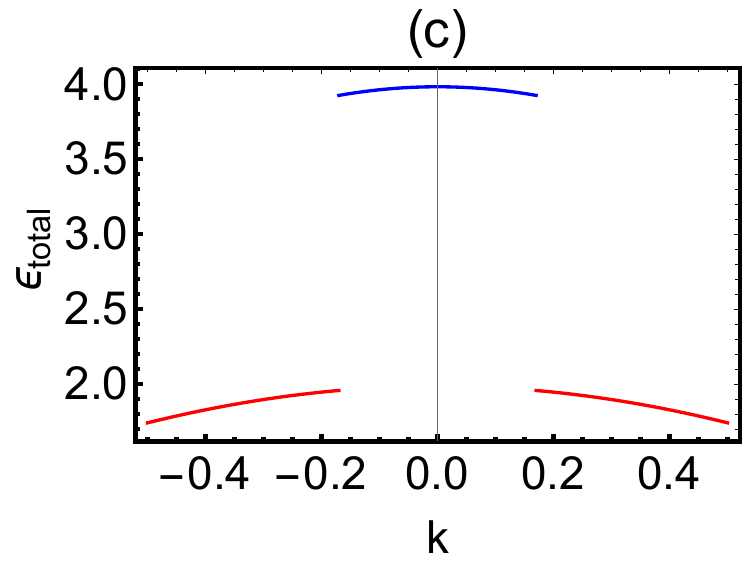}
    \includegraphics[width=0.6\linewidth]{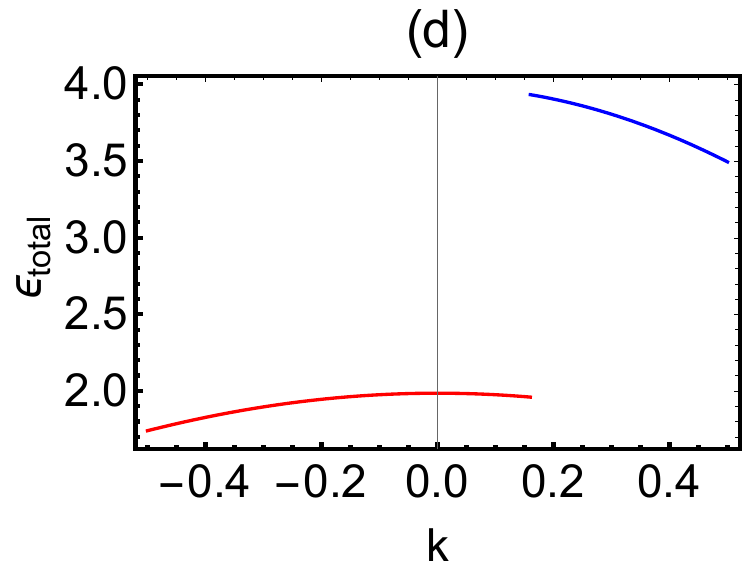}
    \caption{Total energy ($\epsilon_{total}$) -momentum ($k$) dispersion diagram for 1D of (a) BP1 (b) BP2 (c) PS (d) FFLO at $P=0.5$ : Blue and red indicates paired and unpaired region, here momentum in unit of $\pi$. }
    \label{dispersion} 
\end{figure}
If a system has a population imbalance between the up-spin and down-spin fermions, then different exotic phases arise, because the paring cannot entirely take place in the BCS way. The phases that we consider here are
(i) Breached-Pair-1 state (BP1) or Sarma phase  (ii) Breached pair-2 state (BP2) (iii) phase separation (PS), and (iv) FFLO. Fractional imbalance is defined as $P=\frac{n_{\uparrow}-n_{\downarrow}}{n_{\uparrow}+n_{\downarrow}}$. Here, $n_{\uparrow}$ and $n_{\downarrow}$ stand for number of up-spin and down-spin respectively.
\begin{figure}[b]
    \centering
    \includegraphics[width=0.5\linewidth]{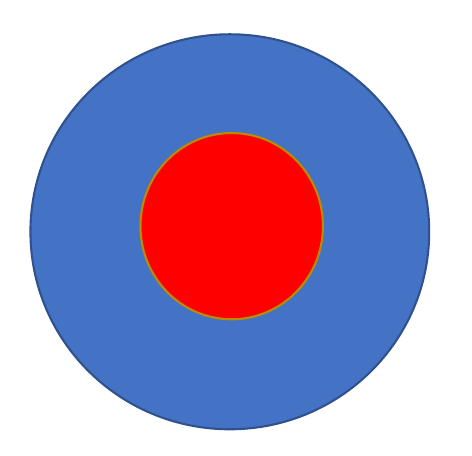}
    \caption{The momentum-space structure of BP1 phase in 2D : the blue shell contains superfluid and the red region contains free fermi gas}
    \label{BP1 2D}
\end{figure}
\begin{figure}[t]
    \centering
    \includegraphics[width=0.5\linewidth]{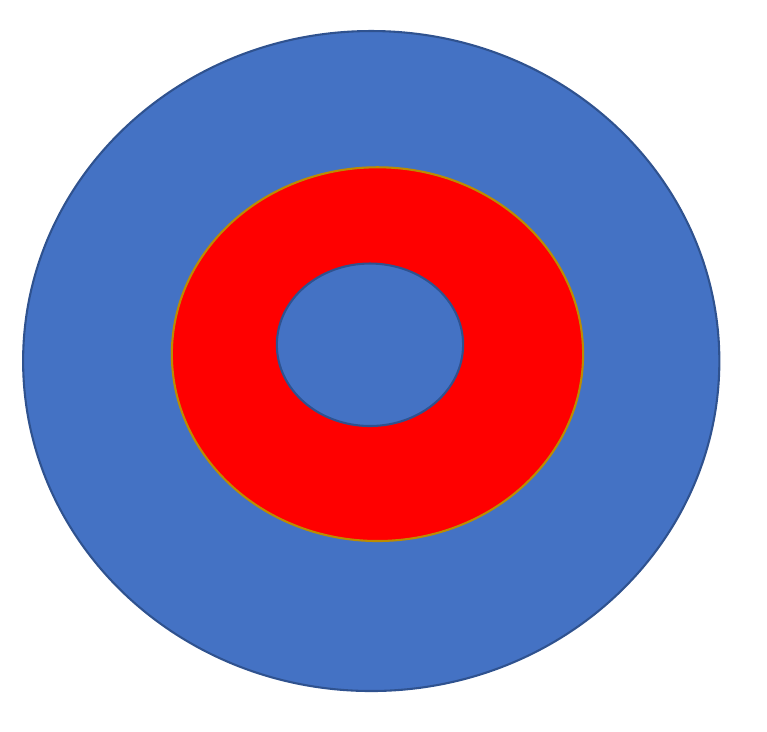}
    \caption{The momentum-space structure of BP2 phase in 2D : the blue regions contains superfluid and the red region contains  free fermi gas}
    \label{BP2 2D}
\end{figure}

In Fig. \ref{all phases}, the blue region and the red dashed regions indicate the paired and unpaired region respectively for a one-dimensional Fermi gas, and different types of pairing structures are depicted. Breached pair or Sarma phase  appears at $T=0$, $\Delta\neq 0$, $P\neq 0$ (here, $\Delta$ is the energy gap). In the BP1 (Fig. \ref{all phases} (a)) structure for a 1-dimensional system,   we get the paired region near the ends of the momentum-space line, and the unpaired free fermi gas remains in the middle. In BP2 (Fig. \ref{all phases} (b)) structure for 1-dimension, we get the paired region at both ends of the line and also in the middle, and the ``breached" regions with a single species exist in-between.  Phase separation (Fig. \ref{all phases} (c))  is defined as the coexistence of superfluid ($\Delta\neq 0$, $P= 0$) and normal states ($\Delta=0$, $P\neq 0$). In Fig. \ref{all phases} (c), the paired region expands about $k=0$.  FFLO (Fig. \ref{all phases} (d))  state is a BCS-like paired state but with a non-zero total momentum. FFLO state is not translationally invariant like BCS and BP phases. Here, the gap parameter is $\Delta(r)=\Delta e^{2i\vec q .\vec r}$ ($q>0$) where $\Vec{r}$ is the distance from the origin. 
\begin{figure}
    \centering
    \includegraphics[width=0.5\linewidth]{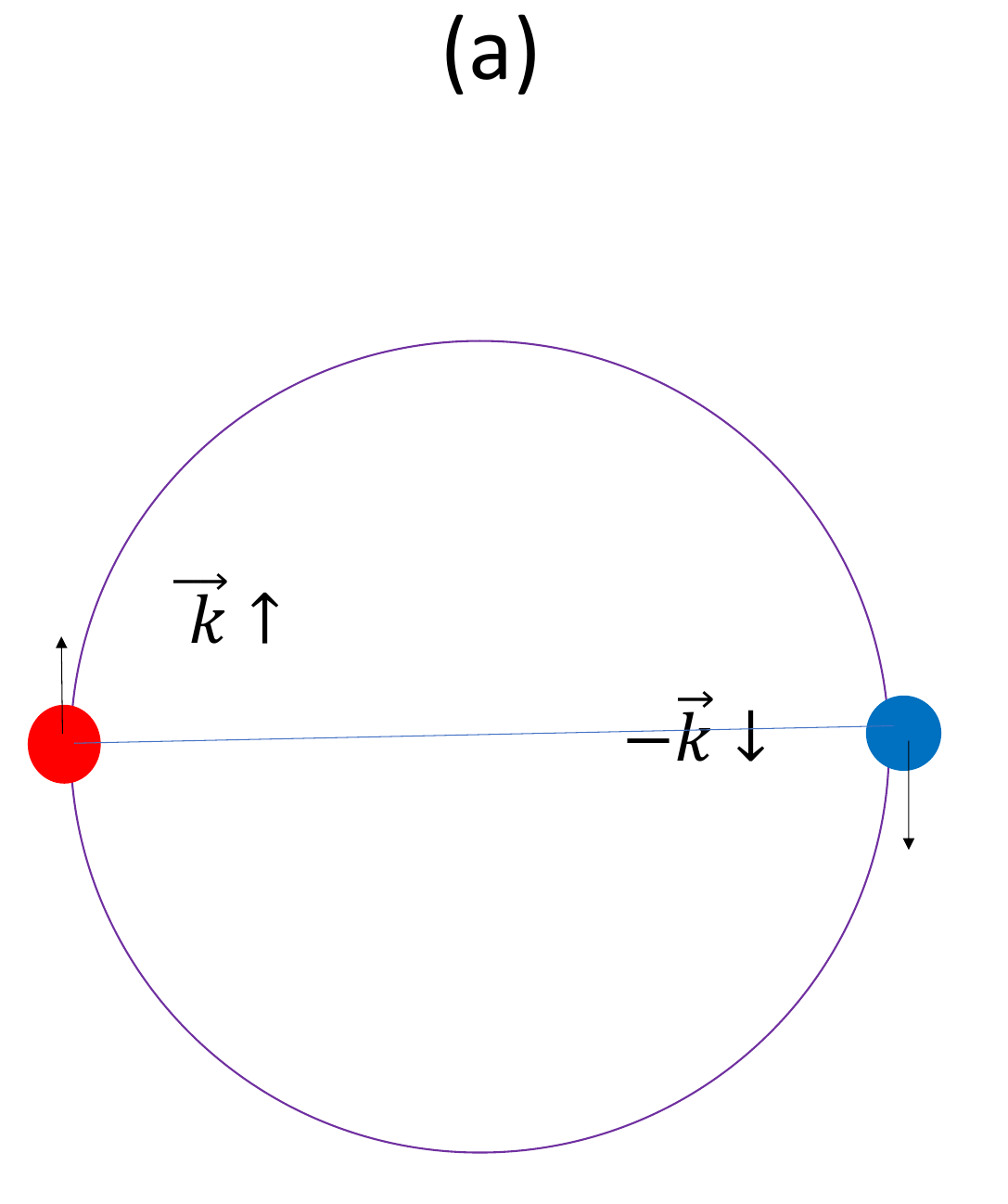}
    \includegraphics[width=0.5\linewidth]{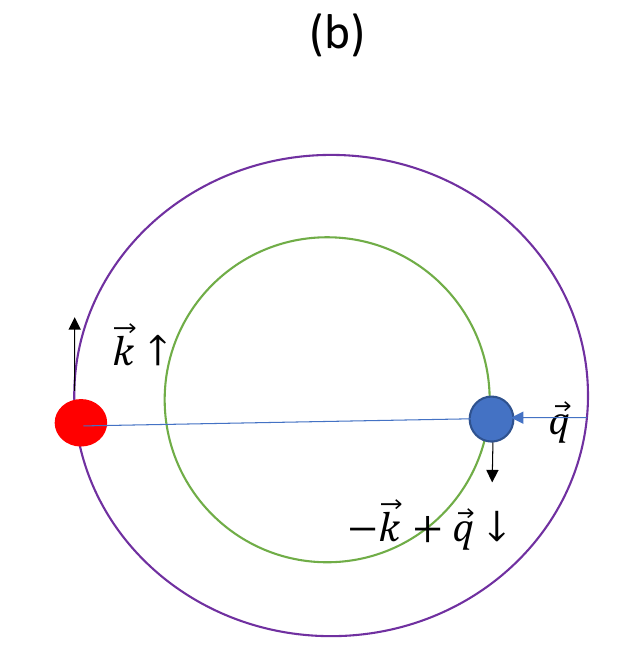}
    \caption{(a) Usual BCS type of pairing (b) FFLO type of pairing}
    \label{FFLO 2D}
\end{figure}
Out of these phases, BP1, BP2 and FFLO involve specific structures in momentum space. Although these phases have been predicted long back, direct experimental signatures of these phases are difficult to obtain. In contrast, phase separation appears in real space, and is easier to identify experimentally \cite{ps_reference3}.
\begin{figure}
  \includegraphics[width=0.5\linewidth]{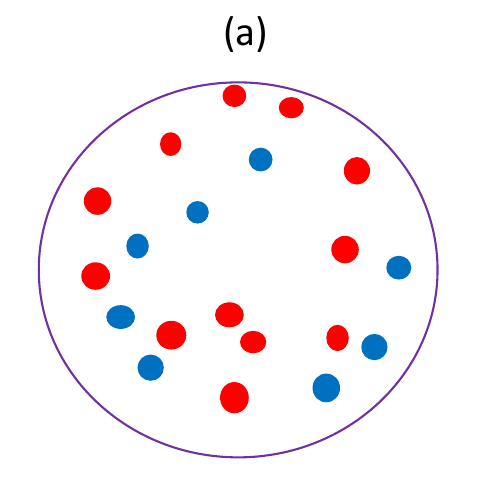}
    \includegraphics[width=0.5\linewidth]{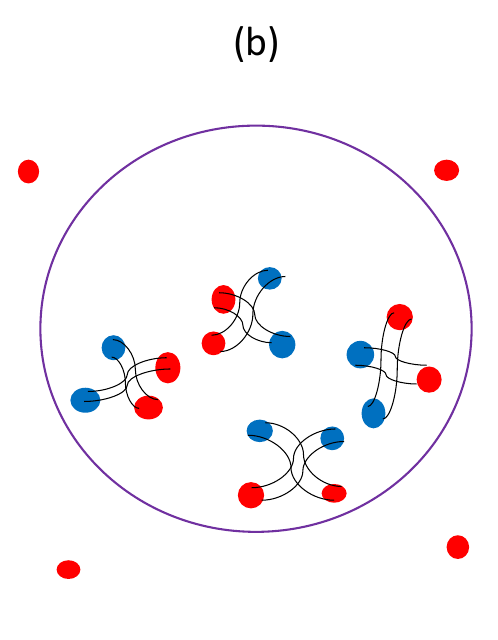}
    \caption{(a) low attraction strength between red and blue atoms (b) moderate attraction strength between red and blue atoms which generate the phase separation state}
    \label{Normal}
\end{figure}
Here the value of $f_1(\omega)$ in Eq. (\ref{freq}) has to be evaluated by using Eq. (\ref{f1 1D}). The limit of the integration is chosen in accordance with the structure of the pairing phases as shown in Fig. \ref{all phases}. As an example, for fractional imbalance, $P=0.5$, we put the limits of the $k$ value of the paired region as $k=-0.5\pi$ to $k=-0.33\pi$ and $k=0.33\pi$ to $k=0.5\pi$ for BP1. The paired region in BP2 is taken as $k=-0.5\pi$ to $k=-0.42\pi$, $k=-0.08\pi$ to $k=0.08\pi$ and $k=0.42\pi$ to $k=0.5\pi$. The paired region in PS spans from $k=-0.17\pi$ to $k=0.17\pi$. In FFLO, the paired region is $k=0.17\pi$ to $k=0.5\pi$. The signature of different pairing phases (BP1, BP2, PS, FFLO) is depicted in terms of total energy ($\epsilon_{total}$) and momentum ($k$) dispersion diagram in Fig. \ref{dispersion}. Here, $\epsilon_{total}$ is the total energy of up-spins and down-spins for the paired region and only up-spins in the unpaired region. In 2D or 3D, the BP1 (Fig. \ref{BP1 2D}) is a two-shell structure of the free fermi gas and the superfluid in momentum space. The BP2 state (Fig. \ref{BP2 2D}) in 2D and 3D is a three-shell structure of the free fermi gas and the superfluid in momentum space.  In  Fig. \ref{FFLO 2D}, the nature of FFLO pairing is depicted where there is the non-zero momentum pairing between red (up-spin) and blue atoms (down-spin) \cite{BCS-FFLO} in 2D. The phase-separated state, on the other hand, originates when the up-spin (red) and down-spin (blue) atoms pair with each other and repel the redundant atoms \cite{ps_diagram}, leading to a phase separation between the paired superfluid fermions and the excess unpaired fermions (as shown in Fig. \ref{Normal}). \\
The integration limits for solving Eqs. (\ref{f1 2D}) and (\ref{f1 3D}) numerically in later sections have been chosen in accordance with the pairing structures described here, and are elaborated further in \ref{4th section}. 
\subsection{System parameters}
In this subsection, we set up the parameter values for different types of paring structures to numerically solve Eq. (\ref{freq}). We choose the following values for the optical lattice depth: $V_x=6 E^F_R$ and $V_y=V_z=40E^F_R$ for 1D; $V_x=V_y=6E^F_R$ and $V_z=40E^F_R$ for 2D; and  $V_x=V_y=V_z=6E^F_R$ for 3D. Here $E^F_R$ is the recoil Fermi energy. The wavelength of the laser beam that creates the optical lattice is chosen to be  $\lambda=825$ nm as in \cite{hopping9_and_lambda}.
In  $x$ direction the hopping amplitude is given by  \cite{hopping1,hopping2,hopping3,hopping4, hopping5,hopping6,hopping7_and_g1,hopping8,hopping9_and_lambda} 
\begin{equation}
  \frac{t}{E^F_R}=\frac{4}{\sqrt{\pi}}\Big(\frac{V_x}{E^F_R}\Big)^{\frac{3}{4}} e^{-2\sqrt{V_x/E^F_R}}  
\end{equation}
 Hopping in $y$ and $z$ directions can be calculated in a similar fashion. The interatomic coupling $g_1$ is given by \cite{hopping7_and_g1}
\begin{equation}
 g_1=\sqrt{\frac{8}{\pi}} k a_{bg} E^F_R \Big(\frac{V_x}{E^F_R}\Big)^\frac{1}{4} \Big(\frac{V_y}{E^F_R}\Big)^\frac{1}{4} \Big(\frac{V_z}{E^F_R}\Big)^\frac{1}{4} 
 \end{equation}

The background scattering length  ($a_{bg}$) for different fermionic species ($^6\mbox{Li}$ and $^{40}\mbox{K}$) are given in Table \ref{building elements of g1}. 
If the optical lattice potential is quite deep ($\geq 5E^F_R$), only the lowest two energy bands remain relevant. In this case, the additional interaction in the form of Feshbach coupling is given by, 
\begin{equation}
\begin{split}
g_2=&\sqrt{\frac{4\pi\hbar^2 a_{bg}\Delta B\mu_{co}}{m_F}}
({\int dx \mathcal{W}^B_x[\mathcal{W}^F_x]^2} {\int dy \mathcal{W}^B_y[\mathcal{W}^F_y]^2}\\ & {\int dz \mathcal{W}^B_z[\mathcal{W}^F_z]^2})
\end{split}
\end{equation}
 Here $\mathcal{W}^B_x$, $\mathcal{W}^B_y$, $\mathcal{W}^B_z$ are Wannier functions for bosons along x, y, z direction respectively, and $\mathcal{W}^F_x$, $\mathcal{W}^F_y$, $\mathcal{W}^F_z$ are their fermionic counterparts \cite{wannier1}. Again, for sufficiently deep lattice potentials, the Wannier functions can be approximated by harmonic oscillator wavefunctions. For  $x$-direction, we thus take the bosonic Wannier function to be 
 \begin{equation} 
 \mathcal{W}^B_x=\Big(\frac{m_B\omega^B_x}{\pi\hbar}\Big)^{\frac{1}{4}}e^{-(m_B\omega^B_x x^2)/2\hbar}
 \end{equation} 
 and fermionic Wannier function to be 
 \begin{equation}
 \mathcal{W}^F_x=\Big(\frac{m_F\omega^F_x }{\pi\hbar}\Big)^{\frac{1}{4}}e^{-(m_F\omega^F_x x^2)/2\hbar}
 \end{equation}
 as in \cite{wannier2}.
Here, $m_B$ and $m_F$ are the masses of the bosonic molecule and fermionic atoms respectively. In addition, we consider $\mu_{co}=2\mu_B$ \cite{source_of_table}, $\omega^B_x=\frac{2}{\hbar}\sqrt{V_x E^B_R}$ \cite{omega}, $\omega^F_x=\frac{2}{\hbar}\sqrt{V_x E^F_R}$ \cite{omega}. Here, $E^B_R(=\frac{\hbar^2 k^2}{2 m_B})$ is the bosonic recoil energy and $\Delta B$ is the resonance width. Similarly, $\mathcal{W}^F_y$ and $\mathcal{W}^B_y$ can be computed from $V_y$; while $\mathcal{W}^F_z$ and $\mathcal{W}^B_z$ can be computed from $V_z$. 

Combining all these, the coupling $g_2$ can be expressed as :
\begin{equation}
\begin{split}
\label{g2}
   g_2=\sqrt{\frac{4\pi a_{bg}\Delta  B\mu_{co}}{m_F\hbar}} \Big(\frac{2m_B}{\pi}\Big)^{\frac{3}{4}}\Big( E^B_R\Big)^{\frac{3}{8}} \Big(V^B_x V^B_y V^B_z\Big)^\frac{1}{8}\\
\Big(\frac{2}{2+\sqrt{\frac{m_B}{m_F}}}\Big)^{\frac{3}{2}}
\end{split}
\end{equation}
\begin{equation}
\begin{split}
\label{detuning}
   \epsilon_b=\mu_{co} (B-B_0)
\end{split}
\end{equation}
Here, $B_0$ is the Feshbach resonance position and $B$ indicate applied field.\\
 The complete list of system parameters for 1D, 2D and 3D that we compute, and use for solving the dynamical equations are given in Tables \ref{system parameters for 1D}, \ref{system parameters for 2D} and \ref{system parameters for 3D} in Appendix \ref{appendix A}.

\section{Variation of frequency with detuning}\label{4th section}
In this section, we present the numerical solutions of Eq. (\ref{freq}) in different dimensions, for different types of pairing structures in a population-imbalanced system. The result is presented in the form of oscillation frequency ($\omega$) of the condensate fraction vs. detuning ($\epsilon_b$) of the Feshbach resonance for two distinct fermionic systems: $^6\mbox{Li}$ and $^{40}\mbox{K}$. 

\subsection{Dynamics in 1-dimensional systems}
\label{dynamics numerical}

The $\omega$ vs. $\epsilon_b$ plots for $^6\mbox{Li}$ near the narrow resonance (resonance width $\Delta B =0.1$ G , resonance position $B_0=543.25$ G) is shown in Fig. \ref{1d,0.5,0.1g}. Here we find that for all values of $\epsilon_b$, (except near $\epsilon_b=0$) there exists a single value of $\omega$. It implies that the dynamics of $b_{\vec q}$ is periodic : 
\begin{equation}
\label{periodic}
b_{\vec q}(t)=b^{eq}_{\vec{q}}+b_1 e^{i\omega t }
\end{equation} 
Therefore, the condensate fraction goes as
\begin{equation}
\label{periodic_cond}
|b_{\vec q}(t)|^2=|b^{eq}_{\vec{q}}|^2+|b_1|^2 +(b^{eq}_{\vec{q}})^\dagger b_1 e^{i\omega t }+b_1^\dagger b^{eq}_{\vec{q}} e^{-i\omega t}
\end{equation} 
So, essentially it would contain a $\cos{\omega t}$
component. Thus, the dynamics of the condensate fraction would be periodic as well.

\begin{figure}
\begin{center}
\includegraphics[width=1\linewidth]{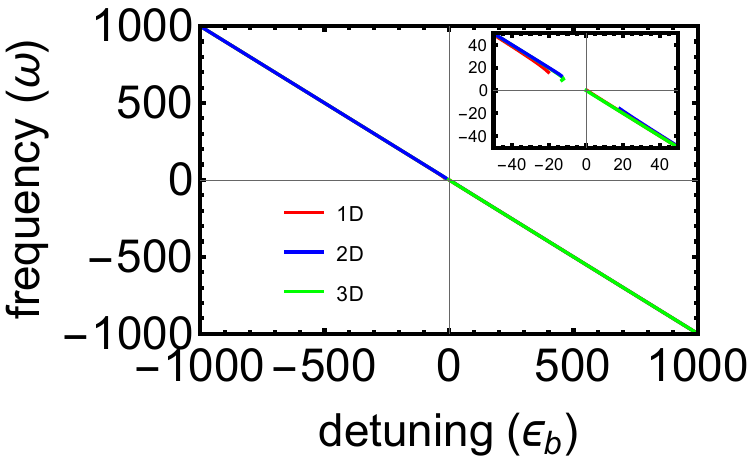}
\caption{$\omega$ vs. $\epsilon_b$ plot at fractional imbalance $P=0.5$ for phase separation of $^6\mbox{Li}$ : (a) Red for 1D (b) Blue for 2D (c) Green for 3D   near narrow resonance. Inset: the enlarged view of the no-oscillation region is indicated by the gap near the resonance point.}
\label{1d,0.5,0.1g}
\end{center}
\end{figure}

The $\omega$ vs. $\epsilon_b$ curve is almost linear in nature. There is a small region with no real solution for $\omega$ in the central region ( Inset of Fig. \ref{1d,0.5,0.1g}). It means that no periodic fluctuation is sustained in the dynamics of the condensate fraction for this particular parameter regime. The plot for $^{40}\mbox{K}$ (resonance width $\Delta B= 7.8$ G, $B_0=202.10$ G) shows a similar pattern in Fig. \ref{1d,0.5,7.8g}, but the region with no solution for $\omega$ is wider. This suggests that the condensate fraction shows no oscillation if the magnitude of the detuning is lower than a threshold value, and there is a periodic oscillation beyond it. The nature of the $\omega$ vs. $\epsilon_b$ plots is similar in all possible pairing phases. 
\begin{figure}
\begin{center}
\includegraphics[width=\linewidth]{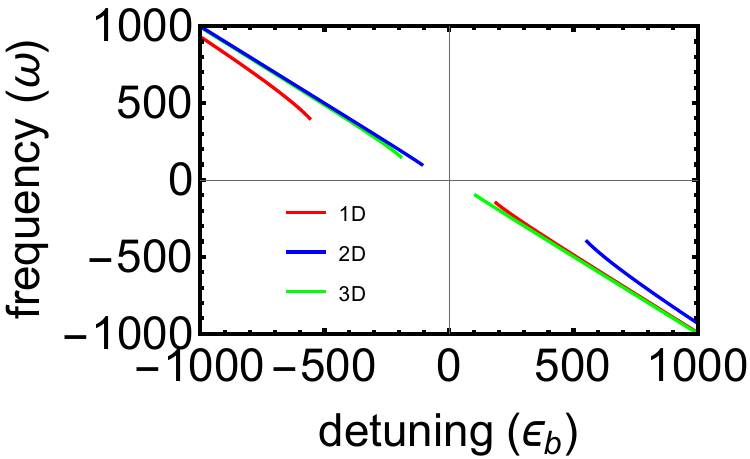}
\end{center}
\caption{$\omega$ vs. $\epsilon_b$ plot at fractional imbalance $P=0.5$ for phase separation in $^{40}\mbox{K}$ (narrow resonance width) : (a) Red for 1D (b) Blue for 2D (c) Green for 3D }
\label{1d,0.5,7.8g}
\end{figure}
We would like to mention that the $\omega$ vs. $\epsilon_b$ curve for the broad resonance in $^6\mbox{Li}$ ($\Delta B= 300$ G, $B_0=834.15$ G) shows similar behaviour. However, the range over which no real solution of $\omega$ is obtained is much wider in this case, and one gets the straight-line-like curve only when the magnitude of the detuning is very large. For the relatively broader resonance of $^{40}\mbox{K}$ ($\Delta B = 9.7 $ G, $B_0=224.21$ G), the $\omega$ vs. $\epsilon_b$ curve is almost similar to the narrower resonance case. These sets of plots (broad resonances of $^6\mbox{Li}$ and $^{40}\mbox{K}$) are not included in the present manuscript. 

\subsection{Dynamics in 2-dimensional systems }

Here, the momentum space phase diagram is a circle. the limits of the integration in Eq. (\ref{f1 2D}) are chosen accordingly.  We set the total radius of the circle as unity, i.e., all other momenta are scaled by the Fermi momentum $k_F$ of the majority species. For fractional imbalance $P=0.5$, we put the limits of the paired region as $k=0.82$ to $1$ for BP1. The paired region in BP2 spans from $k=0$ to $k=0.41$, and $k=0.91$ to $1$. The paired region in PS extends from $k=0$ to $k=0.58$. The paired region of FFLO is $k$=$0$ to $k=0.58$ with shifted origin at $(0.42,0)$. 

The $\omega$ vs. $\epsilon_b$ curve here looks nearly similar to the 1D case. Only the region for no solution is narrower.

\subsection{Dynamics in 3-dimensional systems.}
 The momentum space phase diagram in 3D is a sphere, and we set the total radius of the sphere as unity. As for the limits of the integration in Eq. (\ref{f1 3D}) for fractional imbalance $P=0.5$, the paired region spans from $k=0.87$ to $k=1$ for BP1, from  $k=0$ to $k=0.55$, and $k=0.94$ to $k=1$ for BP2. The paired region spans from $k=0$ to $k=0.69$ for PS. The paired region in FFLO spans from $k$=$0$ to $0.69$ with shifted origin at $\{0.69,0,0\}$.

Here, too,  we find that for all values of $\epsilon_b$, there exists a single value of $\omega$, meaning the oscillation of the condensate fraction is periodic as in Eq. (\ref{periodic}).

The same trend (i.e., a periodic oscillation except near the resonance point) is observed in this case as well. The span of the region of no solution is less than the corresponding regions in 1D.  The $\omega$ vs. $\epsilon_b$ curve is almost linear.

Thus, by comparing  Figs. \ref{1d,0.5,0.1g}, and  \ref{1d,0.5,7.8g} we see that the phase plots do not differ much if the dimension is altered. Also, no major difference vis \`a vis the choice of the atoms ($^6\mbox{Li}$ vs. $^{40}\mbox{k}$) is found. 

We would like to highlight here that the frequencies of oscillation of the condensate fraction fall well within the experimentally detectable range. If the range of detuning is $5$ G to $50$ G \cite{magnetic_field_for_detuning} away from the resonance point ($B_0=543.25$ G \cite{source_of_table} for narrow resonance width of $^6\mbox{Li}$.),  we get the frequency in GHz range for all possible dimensions.  As an example, in Figs. \ref{1d,0.5,0.1g} and \ref{1d,0.5,7.8g} the $\epsilon_b=1000$ point in our scale implies a detuning of $1$ G apart from the resonance point. The corresponding oscillation frequency is $\omega=1000$. In our convention, this translates to a frequency of $0.019$ GHz. In all these numerical computations we set $\hbar=1$.

Additionally, we observe that if the amount of fractional imbalance is changed, then although the qualitative nature of the curves remains the same, some quantitative changes are involved. To quantify these changes, we cast each  $\omega$ versus $\epsilon_b$ plot in the form 
\begin{equation}
\label{linearize}
\omega=b\epsilon_b+c
\end{equation}

 Here $b$ and $c$ are the slopes and intercepts of the line respectively. 

In the next section, we report the effects of population imbalance on the intercept ($c$) and slope ($b$) of this $\omega$ vs. $\epsilon_b$ plots. 
\section{Effects of population imbalance}
The effective scattering length \cite{book1,book2} 
\begin{equation}
\label{scattering length}
    a_s=a_{bg} (1-\frac{\Delta B}{B-B_0})
\end{equation}
In Eq. \ref{scattering length} ${a_s}$, ${a_{bg}}$, ${\Delta B}$, and ${(B-B_0)}$ are scattering length, background scattering length, resonance width and applied magnetic field for $\epsilon_b$ respectively. For the atoms, we consider ($\mbox{Li}^6$ and $\mbox{K}^{40}$) $a_{bg}>0$. It is clear that if $B-B_0>0$ i.e $\epsilon_b>0$, and $\frac{\Delta B}{B-{B_0}}<1$ then $a_s>0$ which implies BEC. If $\epsilon_b<0$, then $a_s>0$ which is also in BEC. We only concentrate on the BEC side for oscillation of condensate fraction.

\subsection{Variation of intercept with fractional imbalance} \label{6th section}
In this section, the linearized form of the $\omega$ vs. $\epsilon_b$ plot is studied and the intercept is extracted. Here we investigate the effect of the population imbalance on the value of this intercept. The result is presented in the form of intercept ($c$) versus fractional imbalance ($P$) plot for $^6\mbox{Li}$. We find that these plots are parabolic in nature. 

  Fig. \ref{1d,0.1g,+ve,intercept} shows that if the fractional imbalance increases, then the absolute value of the intercept decreases for all the phases in $^6\mbox{Li}$ for frequency range $16 \leq \omega \leq 1000$. 

\begin{figure}
\begin{center}
\includegraphics[width=\linewidth]{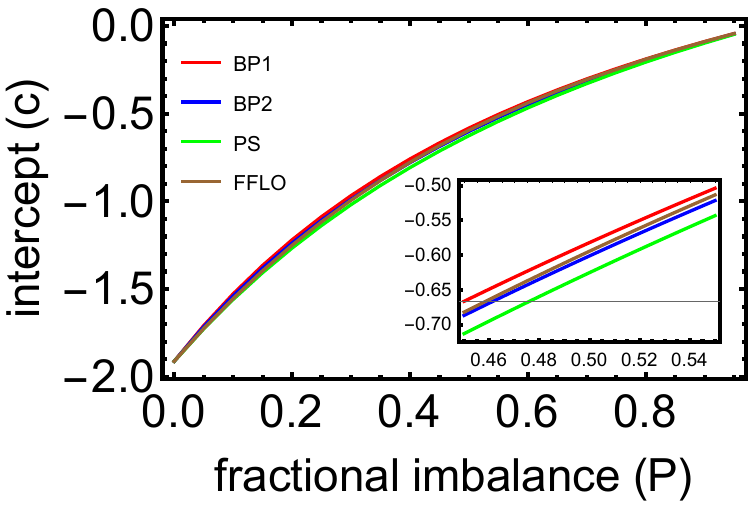}
\caption{$c$ vs. $P$ plot for various phases in positive frequency and a negative detuning domain of $^6\mbox{Li}$ : (a) Red for BP1 (b) Blue for BP2 (c) Green for PS (d) Brown for FFLO at narrow resonance width in 1D. Inset: Distinct signature of four novel phases near $P=0.5$. }
\label{1d,0.1g,+ve,intercept}
\end{center}
\end{figure}

Fig. \ref{2d,0.1g,+ve,intercept}  shows that if the fractional imbalance increases, then the absolute value of the intercept decreases for different phases in $^6\mbox{Li}$ for frequency range $9 \leq \omega \leq 1000$.
\begin{figure}
\begin{center}
\includegraphics[width=\linewidth]{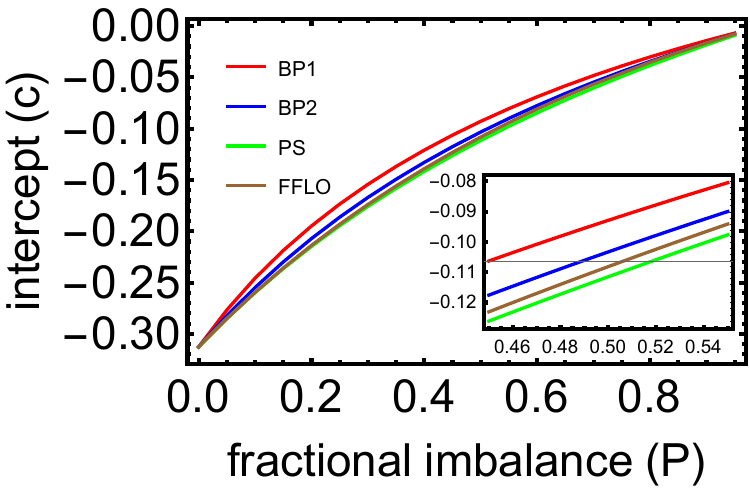}
\caption{$c$ vs. $P$ plot for various phases in positive frequency and a negative detuning domain of $^6\mbox{Li}$ : (a) Red for BP1 (b) Blue for BP2 (c) Green for PS (d) Brown for FFLO  at narrow resonance width in 2D. Inset: Distinct signature of four novel phases near $P=0.5$. }
\label{2d,0.1g,+ve,intercept}
\end{center}
\end{figure}
 Fig. \ref{3d,0.1g,+ve,intercept} shows that like the 1D and 2D cases the $c$ vs. $P$ plot for 3D, too is similar in nature for all possible phases in the frequency range $13 \leq \omega \leq 1000$.  \\
\begin{figure}
\begin{center}
\includegraphics[width=\linewidth]{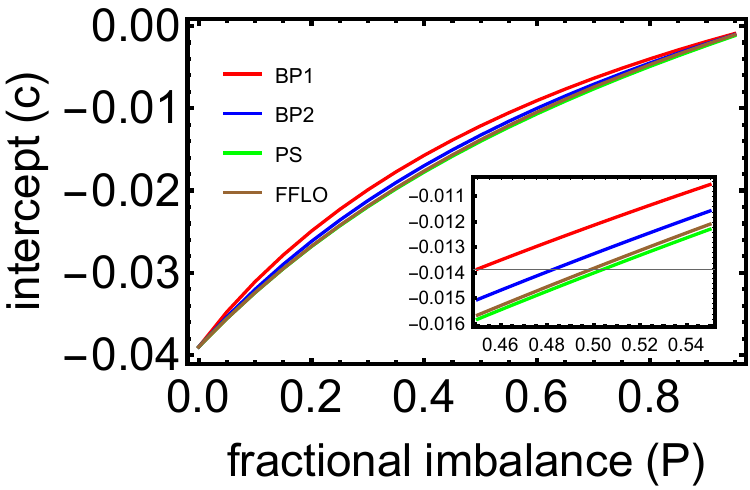}
\caption{$c$ vs. $P$ plot for various phases in positive frequency and a negative detuning domain of $^6\mbox{Li}$ : (a) Red for BP1 (b) Blue for BP2 (c) Green for PS (d) Brown for FFLO  at narrow resonance width in 3D. Inset: Distinct signature of four novel phases near $P=0.5$.}
\label{3d,0.1g,+ve,intercept}
\end{center}
\end{figure}
In the next section, we try to study the effect of $P$ on the slope ($b$) of the linearized form of the $\omega$ vs. $\epsilon_b$ curve.
\subsection{variation of the slope with population imbalance}\label{5th section}

In this section, we study how the $\omega$ vs. $\epsilon_b$ plots in Sec. \ref{4th section} change with a varying amount of population imbalance. For this, we extract the slope $b$ from the linearized form of the curves as in Eq. (\ref{linearize}). The result is presented in the form of a slope ($b$)  versus fractional imbalance ($P$)  plot for $^6\mbox{Li}$. It appears that these curves are parabolic in nature. 
 
 Figs. \ref{1d,0.1g,+ve,slope}, \ref{2d,0.1g,+ve,slope}, \ref{3d,0.1g,+ve,slope} show that if $P$  increases then the absolute value of the slope decreases for four phases in $^6\mbox{Li}$ in 1D ($16 \leq \omega \leq 1000$), 2D ($9 \leq \omega \leq 1000$), and 3D ($13 \leq \omega \leq 1000$). 
\begin{figure}
\begin{center}
\includegraphics[width=\linewidth]{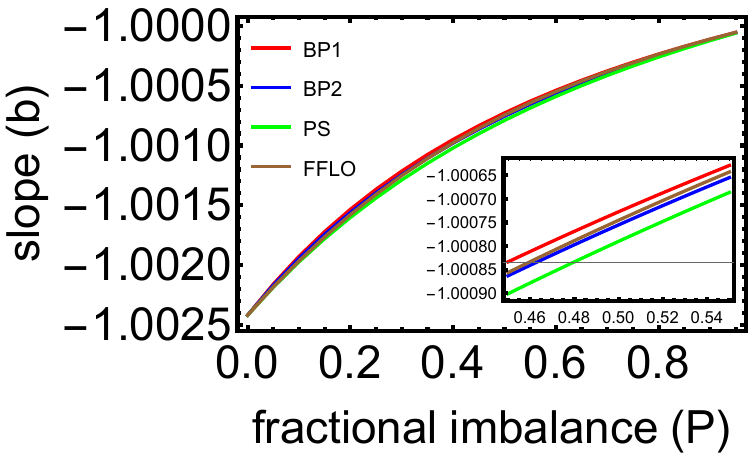}
\caption{$b$ vs. $P$ plots for various phases in positive frequency and a negative detuning domain of $^6\mbox{Li}$ : (a) Red for BP1 (b) Blue for BP2 (c) Green for PS (d) Brown for FFLO  at narrow resonance width in 1D. Inset: Distinct signature of four novel phases near $P=0.5$.}
\label{1d,0.1g,+ve,slope}
\end{center}
\end{figure}
    
\begin{figure}
\begin{center}
\includegraphics[width=\linewidth]{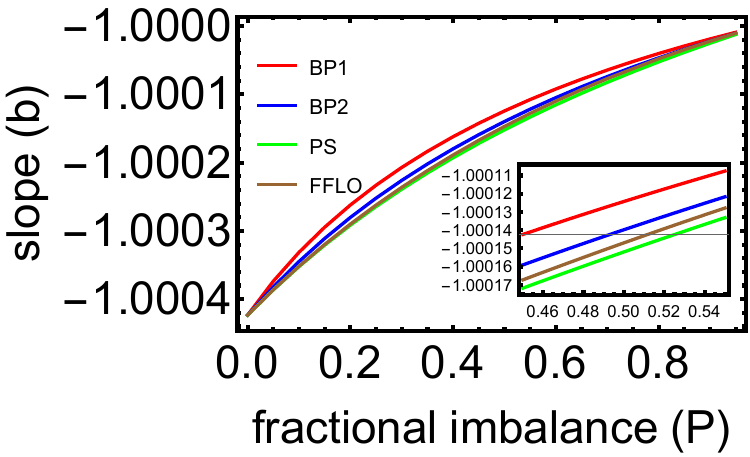}
\caption{$b$ vs. $P$ plot for various phases in positive frequency and a negative detuning domain of $^6\mbox{Li}$: (a) Red for BP1 (b) Blue for BP2 (c) Green for PS (d) Brown for FFLO at narrow resonance width in 2D. Inset: Distinct signature of four novel phases near $P=0.5$.}
\label{2d,0.1g,+ve,slope}
\end{center}
\end{figure}

\begin{figure}
\begin{center}
\includegraphics[width=\linewidth]{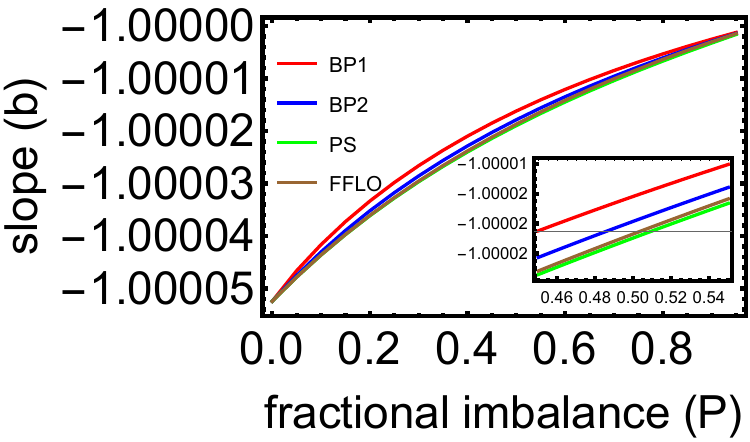}
\caption{$b$ vs. $P$ plot for various phases in positive frequency and a negative detuning domain of $^6\mbox{Li}$ : (a) Red for BP1 (b) Blue for BP2 (c) Green for PS (d) Brown for FFLO of  at narrow resonance width in 3D. Inset: Distinct signature of four novel phases near $P=0.5$.}
\label{3d,0.1g,+ve,slope}
\end{center}
\end{figure}

  It is to be noted that the $P=0$ limit corresponds to perfect BCS-type paiiring with no population imbalance, and the $P=1$ limit corresponds to the presence of a single-fermionic species only. That is why in these two extreme limits, all the phases merge. At the intermediate point ($P=0.5$) the curves corresponding to different phases are maximally separated from one another. 
 
\section{Analyzing the nature of the slopes and the intercepts } \label{7th section}
In this section, we try to explain the results presented in   \ref{6th section} and \ref{5th section}, by means of providing analytical arguments and physical justifications.  We first establish that there would be only one non-trivial solution for $\omega$ from Eq. (\ref{freq}). Then we argue that the absolute value of the slope ($b$)  and the intercept ($c$)  of the $\omega$ vs. $\epsilon_b$ curve depends on (i) the amount of imbalance present, and (ii) the exact nature of pairing. 

\subsection{A single branch of $\omega$ }
We observe that the function $f_1(\omega)$ can be approximated by a simpler function
\begin{equation}
f_1(\omega)\approx \frac{\alpha}{\hbar\omega}.
\label{approx}
\end{equation} 

\begin{figure}
      \centering
      \includegraphics[width=0.4\textwidth]{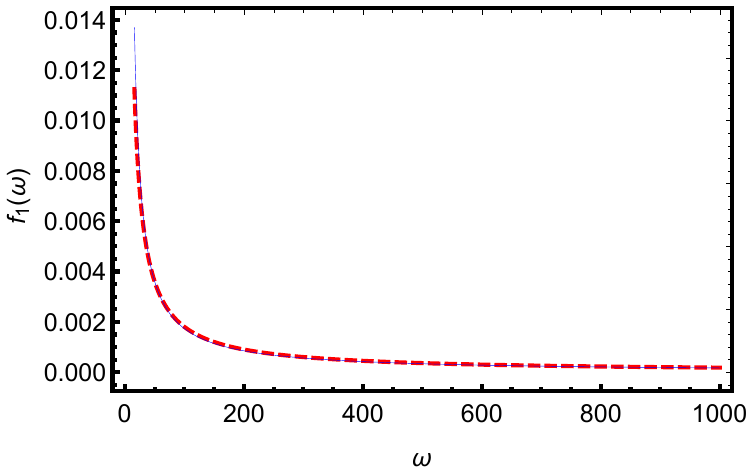}
      \caption{ Plot of $f_1(\omega)$ denoted by blue solid line and its nearest function denoted by red dashed line vs. $\omega$}
      \label{last diagram}
  \end{figure}

Here the value of $\alpha$ depends on (i) the dimension of the system, (ii) the amount of population imbalance present, and (iii) the nature of the pairing. In Fig. \ref{last diagram} we plot $f_1(\omega)$ and $\frac{\alpha}{\hbar\omega}$ for 1D to show how close the two functions are. In this plot, the blue solid line and the red dashed line represent  $f_1(\omega)$ and $\frac{\alpha}{\hbar \omega}$ respectively.  So, Eq. (\ref{freq}) will now be
\begin{equation}
\epsilon_b=-\hbar\omega-\frac{g^2_2 \alpha}{\hbar\omega-g_1 \alpha}
\end{equation}
Simplifying the above equation we  get
\begin{equation}
\label{omega}
\begin{split}
 \omega_{\pm}=\frac{ g_1 \alpha-\epsilon_b}{2\hbar} \pm\frac{\sqrt{(\epsilon_b- g_1 \alpha)^2-4\alpha(g^2_2-g_1\epsilon_b)}}{2\hbar}
 \end{split}
\end{equation}
Thus, we obtain two roots. One root is close to zero and the other root is non-trivial. For positive $\epsilon_b$ values, we have $\omega_{-}>>\omega_{+}$ and $\omega_{+}\approx 0$. Similarly, For negative $\epsilon_b$ values, $\omega_{+}>>\omega_{-}$ and $\omega_{-}\approx 0$. 
Thus, only one significant solution for $\omega$ exists, and that explains the nature of the plots in  Figs. \ref{1d,0.5,0.1g}, and \ref{1d,0.5,7.8g}. Therefore, the condensate fraction's fluctuation is purely periodic as in Eq. (\ref{periodic_cond}).

There exists no real root for $\omega$ if

\begin{equation}
\label{critical detuning}
  \frac{g_1 g_2 \alpha+2\sqrt{\alpha}g^2_2}{g_2+\sqrt{\alpha}g_1} >\epsilon_b>-\frac{g_1 g_2 \alpha+2\sqrt{\alpha}g^2_2}{g_2+\sqrt{\alpha}g_1}
\end{equation}
Then, one gets an imaginary frequency solution. Now, $\Delta \epsilon_b=2(\frac{g_1 g_2 \alpha+2\sqrt{\alpha}g^2_2}{g_2+\sqrt{\alpha}g_1})$, is the range of that detuning which yields imaginary frequency solution. From Table [\ref{imaginary frequency}], we observe that  $\alpha_{PS}>\alpha_{FFLO}>\alpha_{BP2}>\alpha_{BP1}$.  So, we get $(\Delta \epsilon_b)_{PS}>(\Delta \epsilon_b)_{FFLO}>(\Delta \epsilon_b)_{BP2}>(\Delta \epsilon_b)_{BP1}$.

\subsection{Effect of the population imbalance on the slope and intercept of  $\omega$ vs. $\epsilon_b$ curves }\label{slope and intercept} 
The limits of the integration in Eqs. (\ref{f1 1D}), (\ref{f1 2D}) and (\ref{f1 3D}) are controlled by the fractional imbalance $P$. So, $f_1(\omega)$ depends on $P$. If $f_1(\omega)$ is approximated as in Eq. (\ref{approx}), we find that if $P$ decreases, then $\alpha$ increases and vice versa. A set of  $\alpha$ vs. $P$ values is presented in Table [\ref{variation of f1 with fractional imbalance}] for the PS state, and the overall trend remains the same for the other phases as well. 
\subsubsection{Variation of intercept ($c$)  with fractional imbalance ($P$) }\label{intercept}
 \begin{figure}
     \centering
     \includegraphics[width=0.4\textwidth]{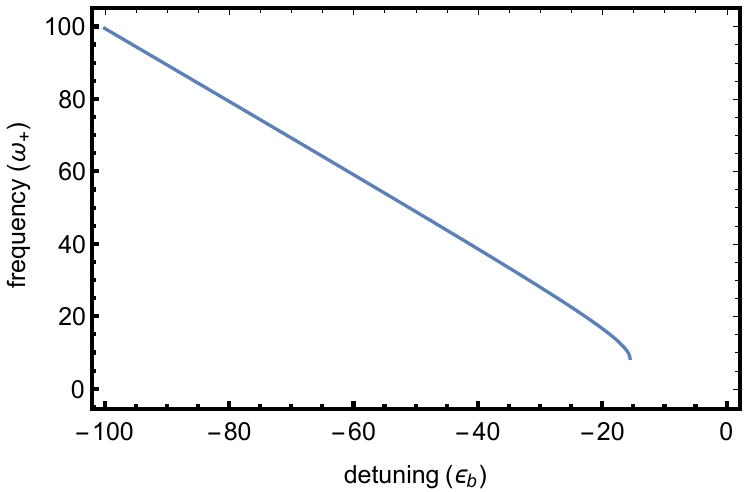}
     \caption{$\omega$ vs. $\epsilon_b$ plot for $^6\mbox{Li}$ (narrow resonance width) at $P=0.5$ for $\epsilon_b>0$ of PS state in 1D.}
     \label{+ve intercept}
 \end{figure}
 The $\omega_+$ vs. $\epsilon_b$ curve for $^6\mbox{Li}$ (narrow resonance width) at $P=0.5$ , as obtained from Eq. (\ref{omega}) is presented in Fig. \ref{+ve intercept}, corresponding to  $\epsilon_b>0$ and PS state in 1D.  It clearly indicates a positive intercept, i.e., $c > 0$. It can be argued that if $\omega_{-}\approx 0$ in Eq. (\ref{omega}), then 
 \begin{equation}
 \label{eqn.26}
    \hbar \omega_{+}\approx -\epsilon_b+g_1 \alpha
\end{equation}
 Thus, $c=g_1\alpha$.
Since if $P$ decreases then $\alpha$ increases, it implies that if $P$ decreases then the intercept $c$ increases, and vice versa.
\subsubsection{Variation of slope ($b$)  with fractional imbalance ($P$)} 

\begin{figure}
      \centering
      \includegraphics[width=0.4\textwidth]{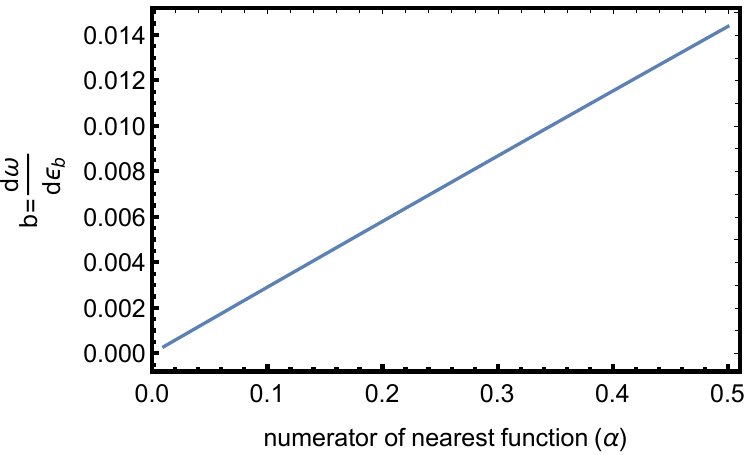}
      \caption{ $b$ vs. $\alpha$ plot for $\epsilon_b<0$}
      \label{slope}
  \end{figure}

The slope of the  $\omega$ vs. $\epsilon_b$ curve, as per Eq. (\ref{omega}) is 
\begin{equation}
\label{omega vs. epsilon_b}
\begin{split}
 b=\frac{d\omega_{\pm}}{d \epsilon_b}=-\frac{1}{2\hbar}  \pm \frac{1}{2\hbar}\frac{\epsilon_b+g_1 \alpha}{\sqrt{(\epsilon_b-\alpha g_1)^2-4\alpha (g^2_2-\epsilon_b g_1)}}
 \end{split}
\end{equation}
We consider the non-trivial branch always, i.e., for $\epsilon_b>0$, $b$ is calculated using $\omega_-$; while for $\epsilon_b<0$, $b$ is calculated using $\omega_+$. Accordingly, $b$ vs. $\alpha$ is plotted in Fig. \ref{slope}. It shows that $b$ increases linearly with $\alpha$ when other parameters are held constant. Thus, if $P$ decreases then $b$  increases. This can explain the traits shown in  Figs. \ref{1d,0.1g,+ve,slope}, \ref{2d,0.1g,+ve,slope},  and \ref{3d,0.1g,+ve,slope}.

  
It is to be noted that Eq. \ref{detuning} and the linearized form Eq. \ref{linearize} imply that the slope ($b$) of the $\omega$ vs. $\epsilon_b$ line accounts for the detuning-dependent (i.e., the magnetic field-dependent) contribution to the oscillation frequency, and the intercept ($c$) accounts for the detuning-independent (i.e., the magnetic field-independent) contribution. Therefore,   factors like (i) the amount of population imbalance present, and (ii) the nature of the pairing are reflected in the value of $c$. This is why the value of $c$ changes appreciably if (i) $P$ is varied, or (ii) one considers different phases but with the same amount of population imbalance. 

On the other hand, $b=\frac{\partial \omega}{\partial \epsilon_b}$ captures the detuning dependence of $\omega$. Therefore, $b$ changes by a minimal amount if either the population imbalance or the pairing phase is altered.

\subsection{Physical significance for order of $b$ and $c$ in various phases}
 We observe that  $ \mid c_{BP1}\mid > \mid c_{BP2}\mid> \mid c_{FFLO}\mid> \mid c_{PS}\mid$  as shown in Figs \ref{1d,0.1g,+ve,intercept} - \ref{3d,0.1g,+ve,intercept}. This is because, when the bosons accumulate near $k=0$, they form a more robust BEC, with less fluctuation. The PS state, which has bosons arranged in a region encircling $k=0$ thus has the lowest $\omega$ value. Between $BP2$ and $BP1$, the $BP1$ phase has more atoms pushed away from the origin in k-space and is the least robust phase among all with the highest $\omega$ value. FFLO state would be somewhere in between, depending on the value of $P$, as its structure is asymmetric and the amount of its skewness is governed by $P$.

 We also observe that and $\mid b_{BP1}\mid >\mid b_{BP2}\mid>\mid b_{FFLO}\mid>\mid b_{PS}\mid$ as shown in Figgs. \ref{1d,0.1g,+ve,slope} - \ref{3d,0.1g,+ve,slope} . This is owing to the fact that when the bosons are concentrated near $k=0$, $\omega$ changes by a small amount even if the detuning is changed. On the other hand, for phases that has bosons occupying higher $k$-values, the change in $\omega$ would be greater for same amount of change in the detuning, as the BEC itself is less stable now. 
 
 The relative change in intercept, as shown in Fig. \ref{3d,0.1g,+ve,intercept} is, however,  $\approx 10 \%$ which can be well captured from experimental data.  From Fig. \ref{3d,0.1g,+ve,slope}, we obtain that the relative variation of the slope across the novel phases is  $\approx 0.001\%$ of the minimum value, which would be difficult to measure in experiments. This, again, is expected because the change in the momentum space structure directly influences $c$, as discussed in [Ref]. Thus, in order to differentiate between the exotic phases effectively, using the intercept data  would be more suitable. 
 
\section{Discussion} \label{8th section}
In this work, we theoretically studied the dynamics of the population-imbalanced ultracold fermions in an optical lattice. Considering a Feshbach coupling that enables the fermionic atoms to form bosonic molecules, we showed that beyond a small threshold value of the Feshbach detuning, the dynamics of the condensate fraction are always periodic. This implies that the oscillation is not just a small ``fluctuation", rather, it is embedded in the mean-field description of the system itself: the periodic dynamics is sustained throughout the course of time. This result is independent of the amount of imbalance present in the system, as well as of the pairing structure. 

In particular, we considered different pairing structures that arise in population-imbalanced fermionic systems, viz., Breached Pair (BP1 and BP2),  Phase Separation (PS), Fulde-Ferrel-Larkin-Ovchinikov (FFLO). Using the exact values of the Feshbach-resonance parameters of $^6\mbox{Li}$ and $^{40}\mbox{K}$, we calculated the oscillation frequencies of the condensate fraction for each of these phases, and in different dimensions. The result is presented in the form of frequency ($\omega$) versus Feshbach detuning ($\epsilon_b$) plots. Again, we observed that neither the nature of the pairing (i.e., whether it is BP1/ BP2 /PS /FFLO) nor the choice of the atoms ( e.g., $^6\mbox{Li}$ or $^{40}\mbox{K}$), dimensionality (1, 2 or 3 dimensions) affects the overall qualitative description: The $\omega$ vs. $\epsilon_b$ curve always shows a straight-line like behaviour. This also implies that the dynamics of the population-imbalanced system in an optical lattice is markedly different from that in a harmonic trap/ homogeneous system \cite{continuous}. 

The equation of the $\omega$ vs. $\epsilon_b$  straight line is dependent on the dimension of the system, the nature of the pairing, and most importantly, the amount of imbalance present. Thus, the slope and intercept of this line can provide interesting insight into the momentum-space structures of the different phases. The amount of imbalance can be deduced for a certain structure from the slope of this line if the slope/ intercept is known for the same system at some other fixed imbalance values. Useful information can also be gathered about the pairing structures, as both the slope and the intercept follow a specific sequence in terms of the exotic phases being present. 

\section{Acknowledgement}
AM would like to acknowledge University Grans Commission (UGC), Govt. of India for financial support (Student ID: 201610064840). RD would like to acknowledge Science and Engineering Research Board (SERB), Dept. of Science and Technology, Govt. of India for providing support under the CRG scheme (CRG/2022/007312).

\section*{Author contribution statement}
Both the authors are involved in all stages of this work. 

\section*{Data availability statement }
All the data obtained from analytical/ numerical calculations have been presented in the forms of graphical plots. Some additional data are presented as tables in the Appendix.

\clearpage\newpage\setcounter{equation}{0} \setcounter{section}{0}
\setcounter{subsection}{0} 
\global\long\def\theequation{S\arabic{equation}}%
\onecolumngrid \setcounter{enumiv}{0} 

\setcounter{equation}{0} \setcounter{section}{0} \setcounter{subsection}{0} \renewcommand{\theequation}{S\arabic{equation}} \onecolumngrid \setcounter{enumiv}{0}
\begin{appendix}
\section {Calculation of parameters for $^6\mbox{Li}$ and $^{40}\mbox{K}$}\label{appendix A}
In this section, we compute the $g_1$ and $g_2$ values for $^6\mbox{Li}$ and $^{40}\mbox{K}$ corresponding to different amounts of fractional imbalance and dimension for  $^6\mbox{Li}$ and $^{40}\mbox{K}$ sample. Building elements of $g_1$ and $g_2$ for both  $^6\mbox{Li}$ and $^{40}\mbox{K}$ has been shown in   Table [\ref{building elements of g1}] in the page of tables.  Table \ref{system parameters for 1D}, Table \ref{system parameters for 2D} and Table \ref{system parameters for 3D} are given on the page of the table for various dimensions of system parameters.
\end{appendix}
\begin{table*}[b]
\caption{Building elements of $g_1$ and $g_2$ \cite{source_of_table}}
\label{building elements of g1}
\begin{tabular}{|l|l|l|l|}
\hline
Species & resonance width ($\Delta B$)  & resonance position ($B_0$) & background scattering length ($a_{bg}$) \\ \hline
$^{6}\mbox{Li}$       & $0.1$ G                 & $543.25$ G             & $59a_0$                           \\ \hline
$^{6}\mbox{Li}$        & $-300$ G                & $834.149$ G            & $-1450a_0$                        \\ \hline
$^{40}\mbox{K}$       & $7.8$ G                & $202.10$ G             & $174a_0$                          \\ \hline
$^{40}\mbox{K}$       & $9.7$ G                & $224.21$ G             & $174a_0$                          \\ \hline
\end{tabular}
\end{table*}

\begin{table*}[b]
\caption{System parameters for 1D}
\label{system parameters for 1D}
\begin{tabular}{|l|l|l|l|l|l|}
\hline
Species & resonance width ($\Delta B$)  & t (set $0.065E^F_r=1$) & $g_1$ (set $0.065E^F_r=1$) & $g_2$ (set $0.065E^F_r=1$) & $\hbar$\\ \hline
$^{6}\mbox{Li}$        & $0.1$ G                 & $1$             & $5.785$  & $17.077$ & 1                        \\ \hline
$^{6}\mbox{Li}$       & $-300$ G                & $1$           & $-137.785$   & $4566.969$  & 1                   \\ \hline
$^{40}\mbox{K}$       & $7.8$ G                & $1$             & $17.062$ & $613.215$ & 1                       \\ \hline
$^{40}\mbox{K}$        & $9.7$ G                & $1$             & $17.062$  & $683.862$ & 1                       \\ \hline
\end{tabular}
\end{table*}
\begin{table*}[b]
\caption{System parameters for 2D}
\label{system parameters for 2D}
\begin{tabular}{|l|l|l|l|l|l|}
\hline
Species & resonance width ($\Delta B$)  & t (set $0.065E^F_r=1$) & $g_1$ (set $0.065E^F_r=1$) & $g_2$ (set $0.065E^F_r=1$)& $\hbar$ \\ \hline
$^{6}\mbox{Li}$        & $0.1$ G                 & $1$             & $3.6$  & $13.477$ & 1                        \\ \hline
$^{6}\mbox{Li}$       & $-300$ G                & $1$           & $-85.738$   & $3602.677$ & 1                    \\ \hline
$^{40}\mbox{K}$        & $7.8$ G                & $1$             & $10.615$ & $483.723$ & 1                        \\ \hline
$^{40}\mbox{K}$       & $9.7$ G                & $1$            & $10.615$  & $539.676$ & 1                       \\ \hline
\end{tabular}
\end{table*}
\begin{table*}[b]
\caption{System parameters for 3D}
\label{system parameters for 3D}
\begin{tabular}{|l|l|l|l|l|l|}
\hline
Species & resonance width ($\Delta B$)  & t (set 0.065$E^F_r=1$) & $g_1$ (set $0.065E^F_r=1$) & $g_2$ (set $0.065E^F_r=1$) & $\hbar$\\ \hline
$^{6}\mbox{Li}$        & $0.1$ G                 & $1$             & $2.246$  & $10.631$ & 1                         \\ \hline
$^{6}\mbox{Li}$       & $-300$ G                & $1$           & $-53.369$   & $2841.6$   & 1                  \\ \hline
$^{40}\mbox{K}$        & $7.8$ G                & $1$             & $6.6$ & $381.738$ & 1                       \\ \hline
$^{40}\mbox{K}$       & $9.7$ G                & $1$             & $6.6$  & $425.538$  & 1                      \\ \hline
\end{tabular}
\end{table*}

\begin{table*}[b]
\caption{Range of detuning ($\Delta \epsilon_b$) which generates an imaginary frequency solution in 1D}
\label{imaginary frequency}
\begin{tabular}{|l|l|l|l|}
\hline
Phase & nearest function ($f_1(\omega)=\frac{\alpha}{\hbar\omega}$)  & numerator of nearest function ($\alpha$) & $\Delta \epsilon_b$ \\ \hline
   PS    & $0.00421729$  & $0.421729$ & $40.372$                      \\ \hline
  FFLO  & $0.00421212$  & $0.421212$  & $40.328$        \\ \hline
  BP2 & $0.004196$   & $0.4196$  & $40.284$              \\ \hline
  BP1 &  $0.00418503$ & $0.418503$ & $40.240$                    \\ \hline
\end{tabular}
\end{table*}

\begin{table*}[b]
\caption{Variation of nearest form of $f_1(\omega)$ with fractional imbalance for PS state and $\mbox{Li}^6$ sample in 1D}
\label{variation of f1 with fractional imbalance}
\begin{tabular}{|l|l|l|l|l|}
\hline
fractional imbalance (P) & paired region & actual form of $f_1$($\omega=16$ to $1000$) & numerator of nearest function( $\alpha$)\\ \hline
$0.1$       &       $-0.41\pi$ to $0.41\pi$ &  $\frac{0.63662 Arc Tanh\frac{-2.9924-0.748591\omega}{15.9999931+(0.005242-\omega)\omega}}{16+(0.005242-\omega)\omega}$       & $0.38$                                    \\ \hline
$0.3$       &    $-0.27\pi$ to $0.27\pi$  & $\frac{0.63662 Arc Tanh\frac{-1.796-0.450\omega}{15.999+(0.0157-\omega)\omega}}{15.999+(0.016-\omega)\omega}$           &    $0.28$                            \\ \hline
$0.5$        &     $-0.17\pi$ to $0.17\pi$ & $\frac{0.63662 Arc Tanh\frac{-1.06829-0.2679\omega}{15.99+(0.0262-\omega)\omega}}{15.9998+(0.0262-\omega)\omega}$          &     $0.18$                                \\ \hline
$0.7$       &      $-0.09\pi$ to $0.09\pi$   & $\frac{0.63662 Arc Tanh\frac{-0.555-0.139\omega}{15.999+(0.036-\omega)\omega}}{15.9997+(0.036-\omega)\omega}$      &      $0.1$                             \\ \hline
$0.9$ & $-0.03\pi$ to $0.03\pi$   &  $\frac{0.63662 Arc Tanh\frac{-0.16449-0.0413603\omega}{15.999+(0.046-\omega)\omega}}{15.9995+(0.046-\omega)\omega}$                 & $0.028$
\\ \hline
\end{tabular}
\end{table*}

\end{document}